\documentclass[runningheads,citeauthoryear]{apinv}
\usepackage{epsfig,cite,graphics}
\usepackage[T2A]{fontenc}
\usepackage[cp1251]{inputenc}
\usepackage{lscape}
\usepackage{longtable}
\usepackage{hyperref}
\usepackage{url}
\usepackage{amsmath}

\begin{document}

\title{New four-color optical photometry of the young stars V521 Cyg and FHO 27
}
\titlerunning{New four-color optical photometry of the young stars V521 Cyg and FHO 27}
\author{Sunay Ibryamov\inst{1} and Evgeni Semkov\inst{2}}
\authorrunning{S. Ibryamov \& E. Semkov}
\tocauthor{S. Ibryamov \& E. Semkov} 
\institute{Department of Physics and Astronomy, Faculty of Natural Sciences, University of Shumen, 115, Universitetska Str., 9700 Shumen, Bulgaria
	\and Institute of Astronomy and National Astronomical Observatory, Bulgarian Academy of Sciences, 72, Tsarigradsko Shose Blvd., 1784 Sofia, Bulgaria
	\newline
	\email{sibryamov@shu.bg}}
\papertype{Submitted on xx.xx.xxxx; Accepted on xx.xx.xxxx}	
\maketitle

\begin{abstract}
Results from new $BVRI$ photometric observations of the pre-MS stars V521 Cyg and FHO 27 collected during the period from February 2015 to July 2019 are presented.
These stars are embedded in the LDN 935, dubbed "The Gulf of Mexico" in the active star-forming complex NGC 7000/IC 5070.
Previous studies revealed them as variables with large amplitude brightness variations.
Our new observations show that the stars continue to exhibit strong photometric variability and fading events in the brightness.
On the basis of the new data received and the data available in the literature, we were able to specify the periodicity in the brightness variations of V521 Cyg.
In the case of FHO 27, we found a significant peak in its periodogram corresponding to 893 days period. 
\end{abstract}


\section*{1. Introduction}

During the pre-main sequence (PMS) phase of evolution, both classes of PMS stars $-$ low-mass (M $\leq$ 2M$_{\odot}$) T Tauri stars and the more massive (2M$_{\odot}$ $\leq$ M $\leq$ 8M$_{\odot}$) Herbig Ae/Be stars show various types of photometric variability (see Hillenbrand et al. 1992; Herbst et al. 1994; Petrov 2003; Reipurth \& Aspin 2010; Audard et al. 2014; Cody et al. 2014).
In some young stars, large amplitude decrease events in their brightness (about 2.8 mag in the $V$ band) are observed.
Stars with such photometric behavior are called UXors named after their prototype UX Orionis (Herbst et al. 1994).
The observed brightness declines usually last from a few days up to some months.
The widespread explanation for the large-amplitude UXors minima involves the obscuration of the central star by the circumstellar disk or from clouds and objects with different sizes and structure; and the variations in the density of dust clumps orbiting at the vicinity of the star.
Detailed studies of the variability of UXors were made by Voshchinnikov (1989), Grinin et al. (1991), Natta \& Whitney (2000), Dullemond et al. (2003), etc.
Herbst \& Shevchenko (1999) proposed the unsteady accretion as an alternative mechanism for the variability of UXors.
In the deepest minima, the color indices of UXors often become bluer. 
This phenomenon is known as "blueing effect" or "color reverse" (see Bibo \& Th\'{e} 1990).

The stars from our study V521 Cyg and FHO 27 are embedded in the dense molecular cloud LDN 935 (Lynds 1962), dubbed "The Gulf of Mexico".
The Gulf of Mexico is located in the active star-forming complex NGC 7000/IC 5070 and contains many young stellar objects as emission line stars, flare stars, T Tauri variables, and Herbig Ae/Be stars.
Results from recent studies of objects in the complex NGC 7000/IC 5070 have been published in Armond et al. (2011), Findeisen et al. (2013), Bally et al. (2014), Poljan\v{c}i\'{c} Beljan et al. (2014), Ibryamov et al. (2015a), Semkov et al. (2017), Giannini et al. (2018), Ibryamov et al. (2018a), Ibryamov et al. (2018b), Froebrich et al. (2018), Bhardwaj et al. (2019), etc.

Our goals in the current study include: performing $BVRI$ optical observations of V521 Cyg and FHO 27; reducing the obtained new data; building the long-term light curves of the objects by using the new data and the ones available in the literature; discussing the photometric behavior of the objects and the causes of their variability; and searching periodicity in the brightness variations of the objects.

Section 2 in the present paper gives information about the process of performing photometric observations and data reduction.
Section 3 describes the obtained results and their interpretation.
Conclusion remarks are provided in Section 4.

\section*{2. Observations}

The $BVRI$ observations of V521 Cyg were performed in the period November 2015$-$July 2019, while the observations of FHO 27 cover the period February 2015$-$July 2019.
The observations were obtained with the 2-m Ritchey-Chr\'{e}tien-Coud\'{e} (RCC) and the 50/70-cm Schmidt telescopes of the Rozhen National Astronomical Observatory (Bulgaria).
Three different types of CCD cameras were used to perform of the observations $-$ VersArray 1300B (1340 $\times$ 1300 pixels, 20 $\times$ 20 $\mu$m pixel$^{-1}$ size and 0.26$\arcsec$ pixel$^{-1}$) and Andor iKon-L (2048 $\times$ 2048 pixels, 13.5 $\times$ 13.5 $\mu$m pixel$^{-1}$ size and 0.17$\arcsec$ pixel$^{-1}$) on the 2-m RCC telescope, and FLI PL16803 (4096 $\times$ 4096 pixels, 9 $\times$ 9 $\mu$m pixel$^{-1}$ size and 1.08$\arcsec$ pixel$^{-1}$) on the 50/70-cm Schmidt telescope.

All frames were taken through a standard Johnson-Cousins set of filters.
Several frames in each filter were obtained each observational night to estimate the brightness of the objects.
Views of the observed field by different optical instruments used are plotted in Fig. 1.
Twilight flat fields in each filter were obtained every clear evening.
All frames obtained with the VersArray 1300B and Andor iKon-L cameras are bias frame subtracted and flat field corrected.
CCD frames obtained with the FLI PL16803 camera are dark frame subtracted and flat field corrected.

The photometric data were reduced using an \textsc{idl} based \textsc{daophot} subroutine.
All data were analyzed using the same aperture, which was chosen to have a 4$\arcsec$ radius, while the background annulus was taken from 9$\arcsec$ to 14$\arcsec$.
As a reference sequence, we used the $BVRI$ comparisons reported in Semkov et al. (2010).
The typical value of errors in the reported magnitudes is 0.01$-$0.02 mag for the $I$ and $R$ bands data, and 0.01$-$0.03 mag for the $V$ and $B$ bands data.

\begin{figure}[h!]
   \centering
   \includegraphics[width=6.19cm, angle=0]{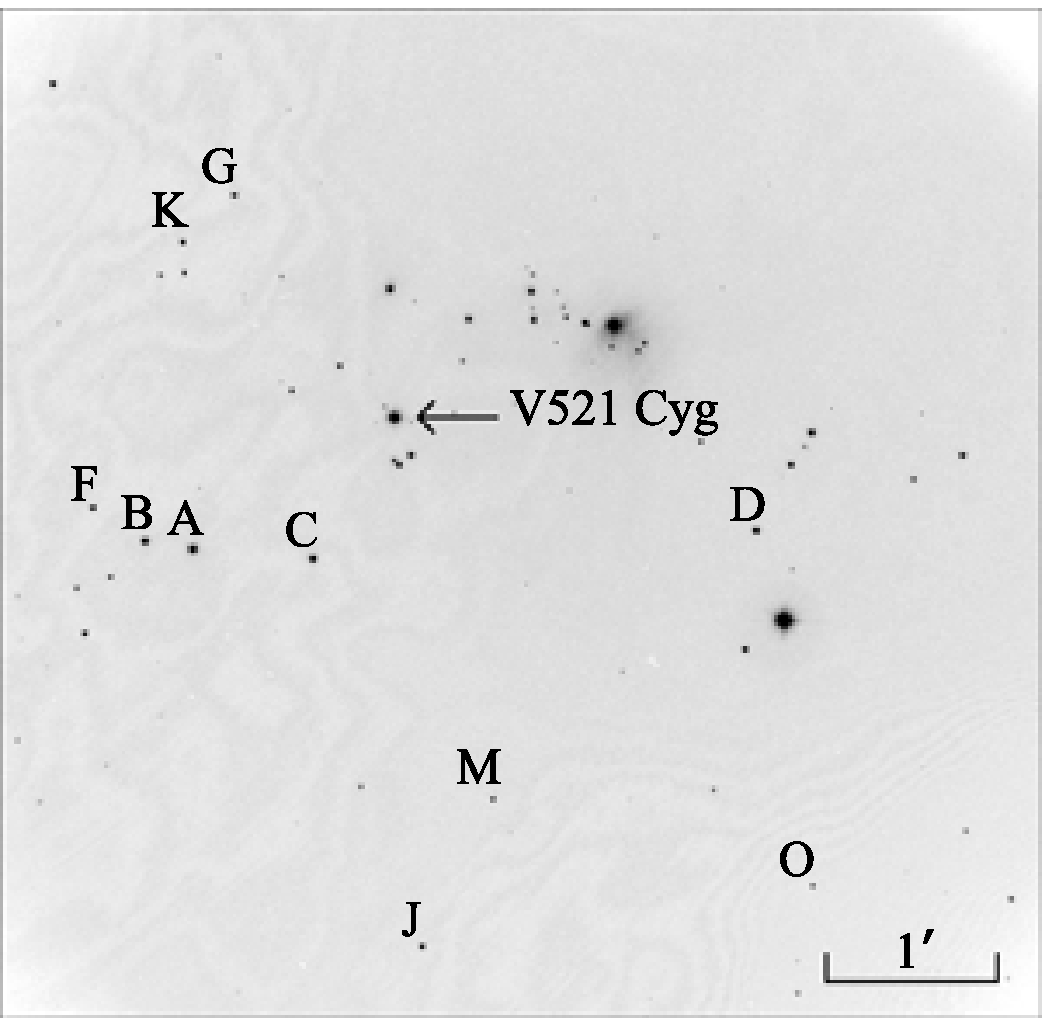}
	 \includegraphics[width=6.00cm, angle=0]{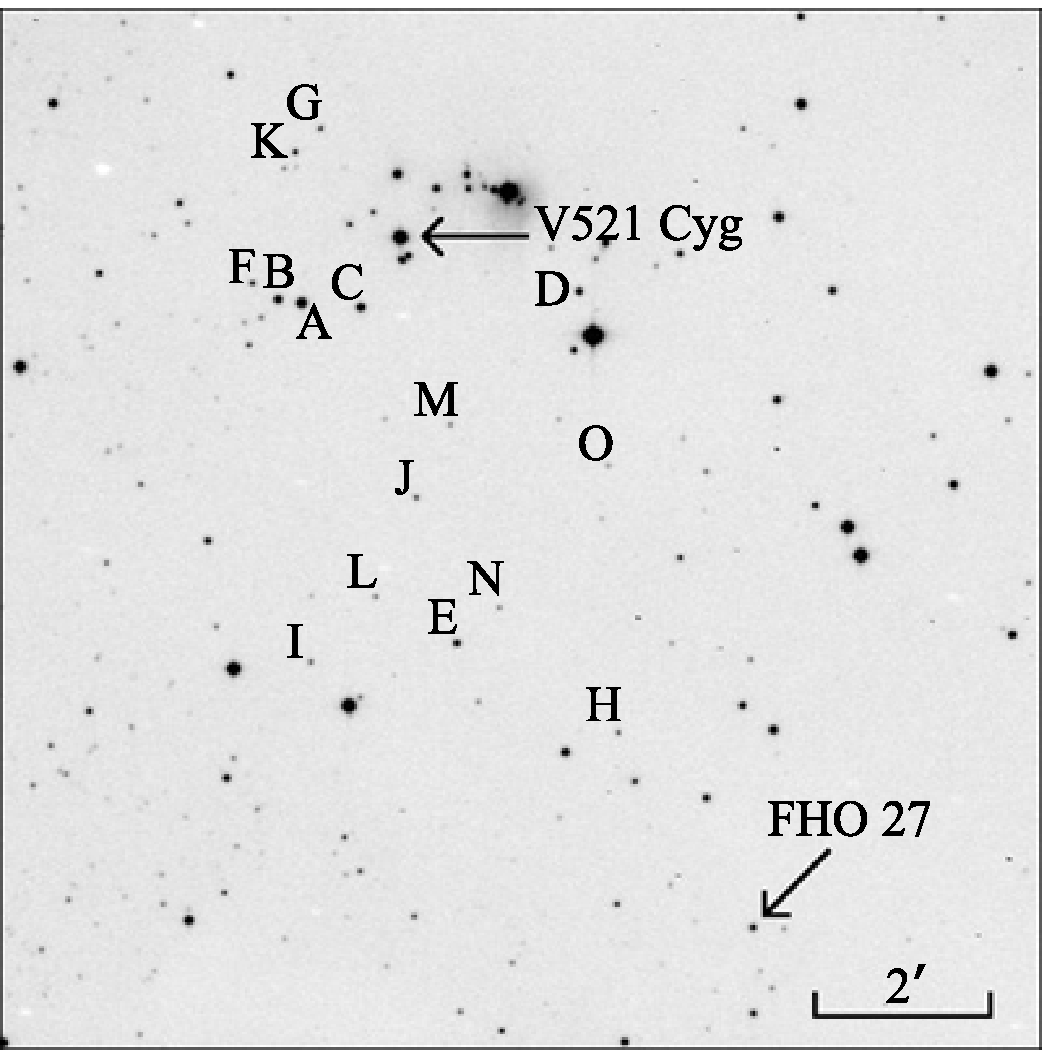}
   \caption{Views of the observed field by different instruments used. North is up, east is left. Left: A frame from the 2-m RCC telescope. The field of view of the 2-m RCC telescope equipped with VersArray 1300B camera (5.76$\arcmin$ $\times$ 5.59$\arcmin$) is almost the same as with Andor iKon-L camera (5.94$\arcmin$ $\times$ 5.94$\arcmin$); Right: A frame from the 50/70-cm Schmidt telescope equipped with FLI PL16803 camera (36.90$\arcmin$ $\times$ 36.90$\arcmin$ by using subframe). For better visualization, the frame from the Schmidt telescope was cropped. The objects from our study and the standard stars from Semkov et al. (2010) are marked.}\label{Fig1}
   \end{figure}

\section*{3. Results and discussion}

\subsection*{3.1. V521 Cyg}

Variability of V521 Cyg was discovered by Hoffmeister (1949).
The star was included in the list of emission-H$\alpha$ stars in the region of NGC 7000 and IC 5070 (Herbig 1958).
Several optical photometric studies were dedicated to V521 Cyg (Filin 1974; Erastova \& Tsvetkov 1978; Mitskevich \& Pavlenko 2001).
They described the star's photometric behavior during different time periods. 
Grankin et al. (2007) analyzed the photometric data of the star collected in the period June 1986$-$September 2003 within the framework of the ROTOR-program at the Maidanak Observatory in Uzbekistan.
The authors reported that the object exhibits unusual color behavior with a blue turnaround at minimum brightness.
This color behavior likely points to variable circumstellar extinction (Grankin et al. 2007).

The results from our $BVRI$ observations of V521 Cyg have summarized in Tab. 1\footnote{The photometric data are available also at CDS (\url{http://cds.u-strasbg.fr/}}).
The columns contain Date (dd.mm.yyyy format) and Julian date of the observations, the measured magnitudes of the star, the telescope and CCD camera used. 
The received new photometric data of the star and the data from Grankin et al. (2007), Poljan\v{c}i\'{c} Beljan et al. (2014), Ibryamov et al. (2015b) and Ibryamov \& Semkov (2016) are presented in Fig. 2. 
In Grankin et al. (2007) the magnitudes of V521 Cyg were provided in the Johnson system.
We used the transformation described in Fernie (1983) to get the values in the $R$ band in the Cousins system.

\begin{figure}[h!]
   \centering
   \includegraphics[width=11cm, angle=0]{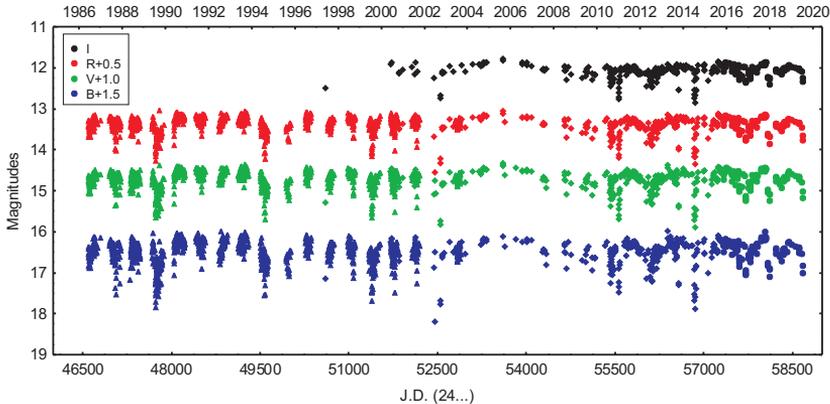}
   \caption{$BVRI$ light curves of V521 Cyg for the period June 1986$-$July 2019. The circles denote the new photometric data (present paper), the triangles represent the data from Grankin et al. (2007), and the diamonds mark the data from Poljan\v{c}i\'{c} Beljan et al. (2014), Ibryamov et al. (2015b), and Ibryamov \& Semkov (2016). The registered amplitudes of the star's variability during the new observations are 0.55 mag in the $I$ band, 0.70 mag in the $R$ band, 0.80 mag in the $V$ band, and 1.11 in the $B$ band.}\label{Fig2}
   \end{figure}

It can be seen from Fig. 2 that from 1986 to mid-2019 there is no change in the magnitude of the star during maximal brightness, and its photometric behavior is characterized by multiple-fading events.
During the period November 2015$-$July 2019, we registered few minima in the light curves of V521 Cyg, which have different amplitudes and durations.
For the same time period, the deepest minimum with amplitudes $\Delta I$ = 0.55 mag, $\Delta R$ = 0.70 mag, $\Delta V$ = 0.80 mag, and $\Delta B$ = 1.11 mag was registered in December 2017.

Using all of our observations we constructed three color$-$magnitude diagrams of V521 Cyg which are displayed in Fig. 3.
It can be seen that the star becomes redder when fainter.
The star becomes redder when its light is being obscured by dust clumps with respect to the line of sight.
In the color indices, there is no blue turning at minimum brightness.
Probably, the obscuration does not rise sufficiently for the scattered light to become considerable enough for registration of color reverse.

\begin{figure}
\begin{center}
\includegraphics[width=7.5cm]{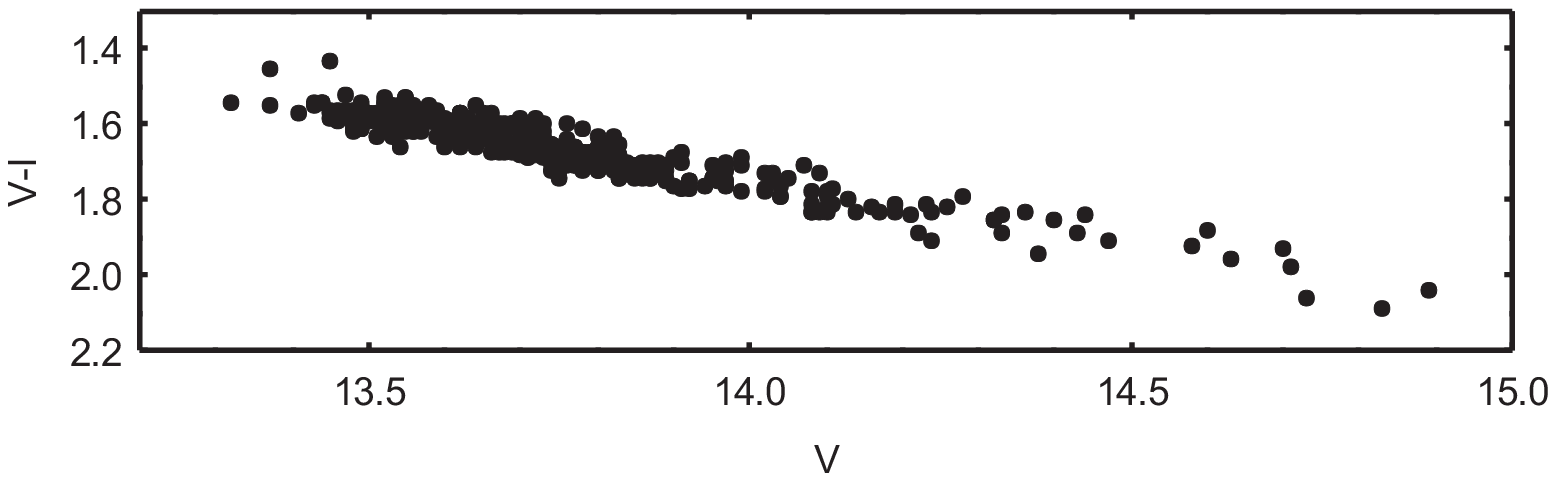}
\includegraphics[width=7.5cm]{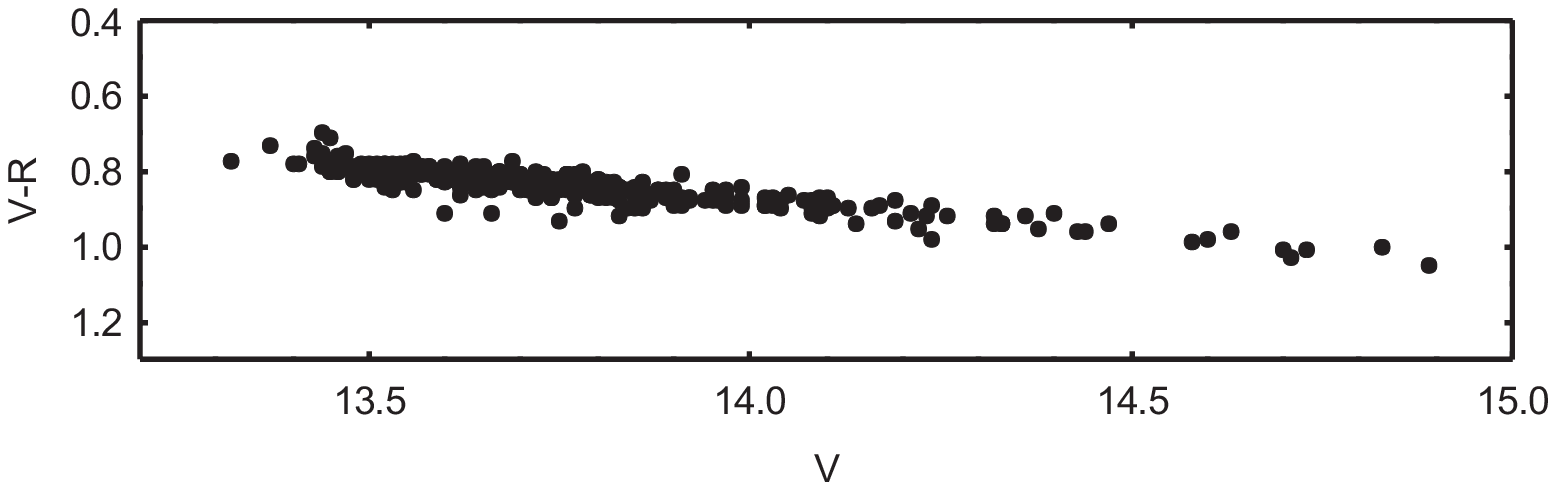}
\includegraphics[width=7.5cm]{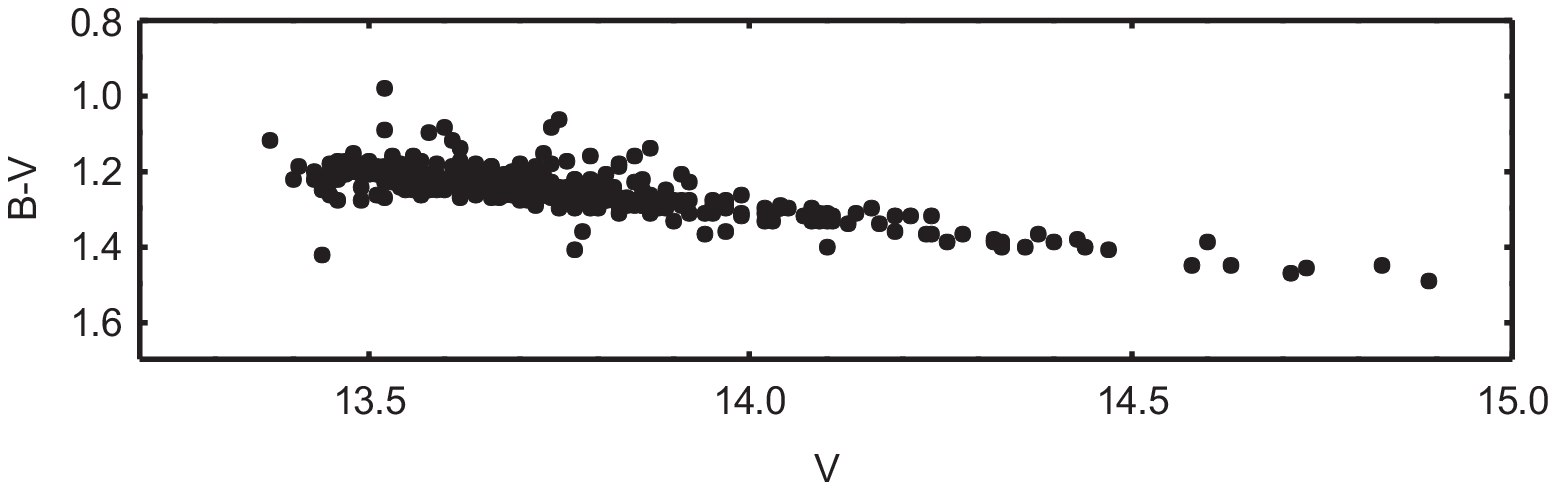}
\caption{Color indices $V-I$, $V-R$, and $B-V$ versus the stellar $V$ magnitude of V521 Cyg.}\label{Fig3}
\end{center}
\end{figure}

On the basis of the received new photometric data and the data published in Grankin et al. (2007), Poljan\v{c}i\'{c} Beljan et al. (2014), Ibryamov et al. (2015b), and Ibryamov \& Semkov (2016), we specified the periodicity in the brightness variations of V521 Cyg.
In the work of Poljan\v{c}i\'{c} Beljan et al. (2014) a period of 503 days was detected.
Our time-series analysis for a periodicity search was conducted with the software package \textsc{period04} (Lenz \& Breger 2005).
The obtained periodogram of the star using the data in the $V$ band is shown in Fig. 4 (left).
We detected a significant peak corresponding to 594 days period.
The $V$ band phased light curve of V521 Cyg according to this period is plotted in Fig. 4 (right).
There are several possible reasons for the periodicity of the star $-$ obscuration by a cloud of gas and dust orbiting at the vicinity of the star, obscuration from a second component or planetesimals, and the precession of the circumstellar disk.

\begin{figure}[h!]
\begin{center}
\includegraphics[width=6cm, angle=0]{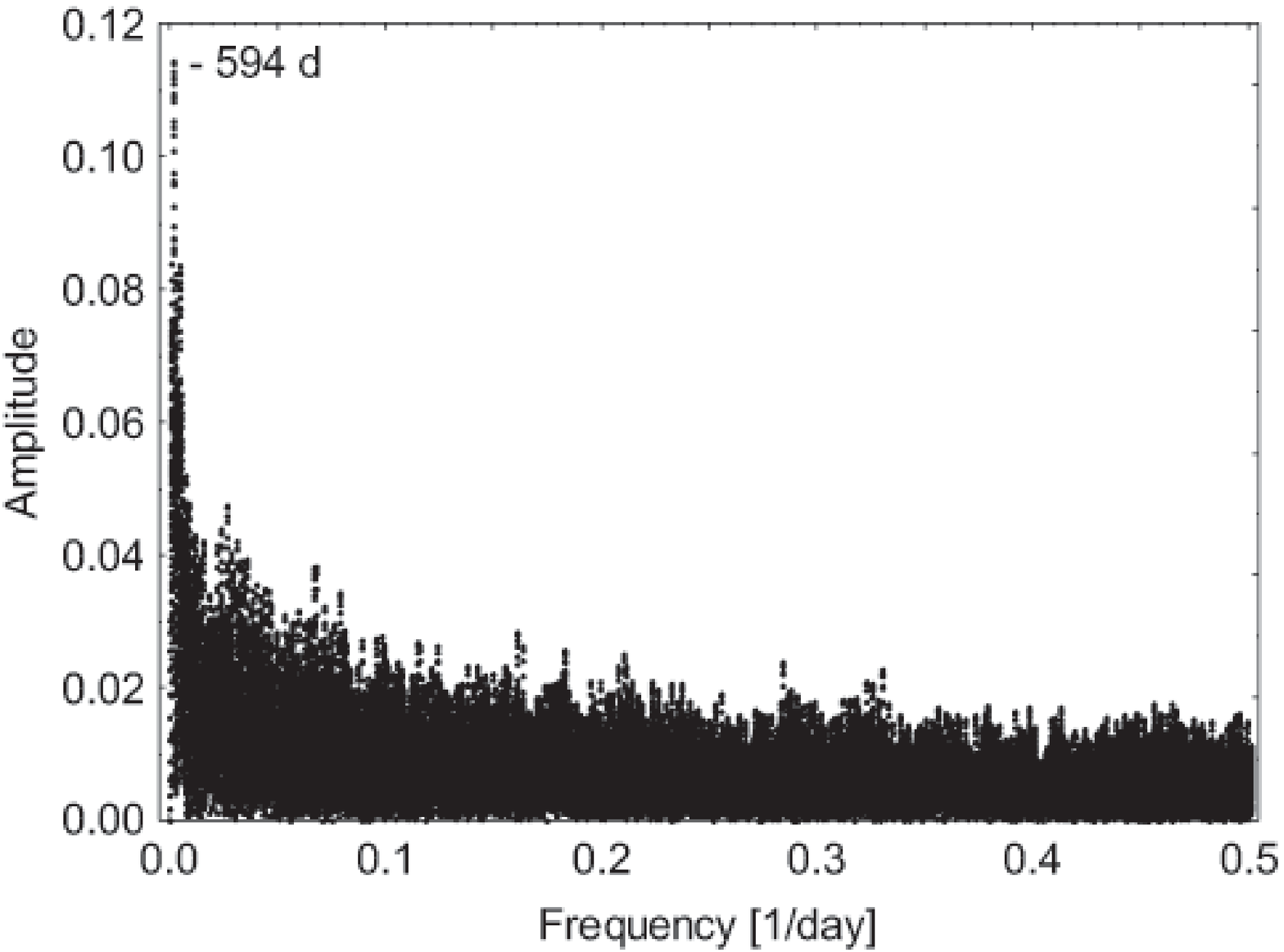}
\includegraphics[width=6cm, angle=0]{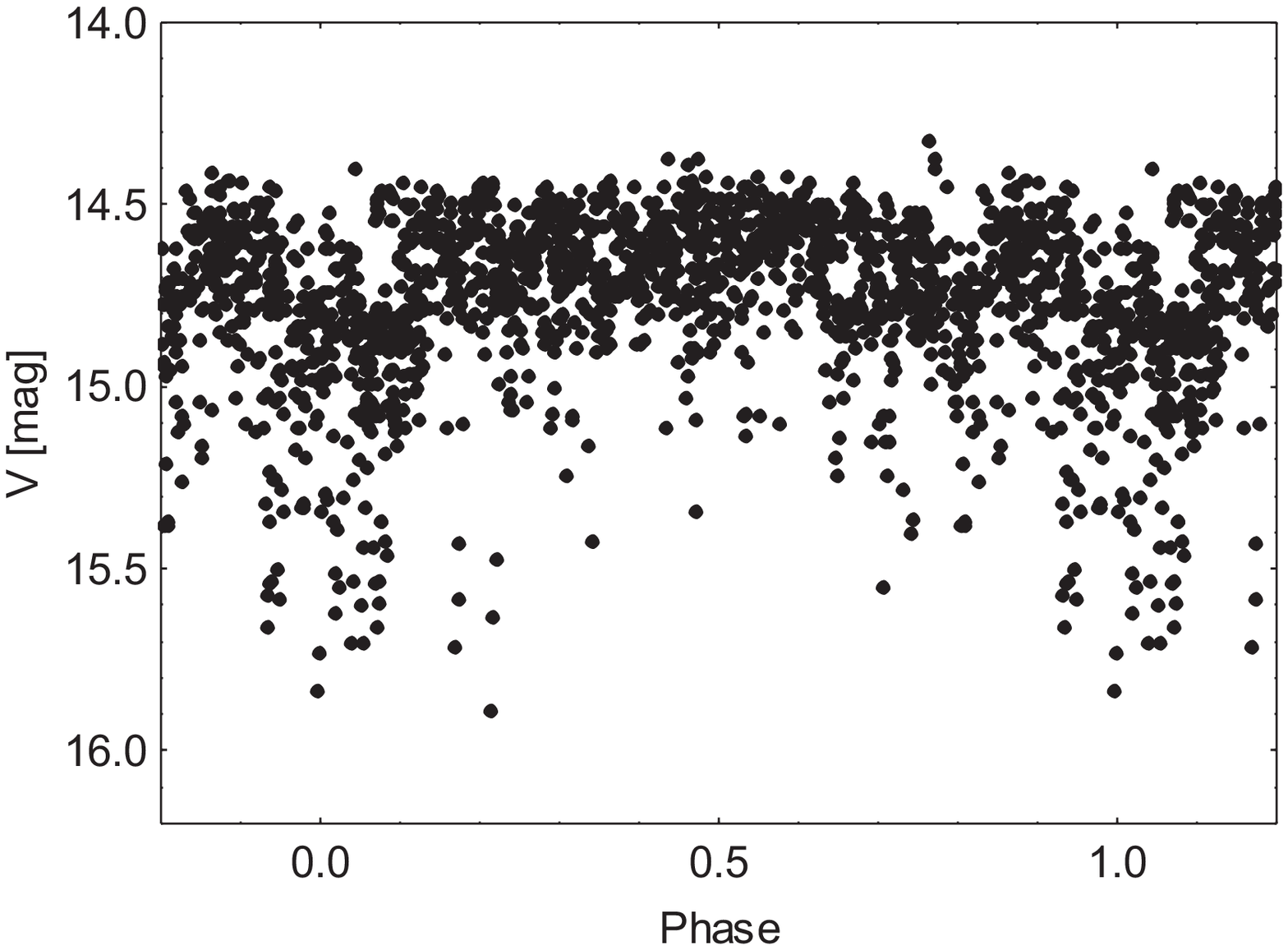}
\caption{Left: Periodogram analysis of the photometric data of V521 Cyg. Right: $V$ band phased light curve of V521 Cyg.}\label{Fig4}
\end{center}
\end{figure}

{\footnotesize
\begin{longtable}{cccccccc}
\caption{$BVRI$ photometry of V521 Cyg}\\
\hline\hline
\noalign{\smallskip}  
Date \hspace{1.0cm} &	J.D. (24...) \hspace{1mm}	&	$I$ [mag]\hspace{2mm} & $R$ [mag]\hspace{2mm} & $V$ [mag]\hspace{2mm} & $B$ [mag]\hspace{2mm} & Tel \hspace{1.0mm} & CCD\\
\noalign{\smallskip}  
\hline
\endfirsthead
\caption{continued.}\\
\hline\hline
\noalign{\smallskip}  
Date \hspace{1.0cm} &	J.D. (24...) \hspace{1mm}	&	$I$ [mag]\hspace{2mm} & $R$ [mag]\hspace{2mm} & $V$ [mag]\hspace{2mm} & $B$ [mag]\hspace{2mm} & Tel \hspace{1.0mm} & CCD\\
\noalign{\smallskip}  
\hline
\noalign{\smallskip}  
\endhead
\hline
\label{Tab1}
\endfoot
\noalign{\smallskip}
03.11.2015	&	57330.191	&	12.09	&	12.92	&	13.74	&	15.00	&	Sch	&	FLI	\\
04.11.2015	&	57331.230	&	12.00	&	12.82	&	13.62	&	14.84	&	Sch	&	FLI	\\
05.11.2015	&	57332.243	&	11.95	&	12.78	&	13.57	&	14.81	&	Sch	&	FLI	\\
06.11.2015	&	57333.228	&	11.99	&	12.79	&	13.58	&	14.83	&	Sch	&	FLI	\\
07.11.2015	&	57334.219	&	11.95	&	12.74	&	13.52	&	14.74	&	Sch	&	FLI	\\
12.12.2015	&	57369.212	&	11.98	&	12.72	&	13.54	&	14.76	&	2-m	&	VA	\\
13.12.2015	&	57370.188	&	11.90	&	12.68	&	13.53	&	14.69	&	2-m	&	VA	\\
14.12.2015	&	57371.199	&	11.86	&	12.66	&	13.48	&	14.68	&	2-m	&	VA	\\
15.12.2015	&	57372.216	&	11.92	&	12.70	&	13.52	&	14.73	&	Sch	&	FLI	\\
17.12.2015	&	57374.269	&	11.90	&	12.70	&	13.46	&	14.74	&	Sch	&	FLI	\\
02.01.2016	&	57390.194	&	11.90	&	12.74	&	13.44	&	14.86	&	Sch	&	FLI	\\
06.02.2016	&	57425.204	&	11.90	&	12.71	&	13.52	&	14.72	&	Sch	&	FLI	\\
07.02.2016	&	57426.200	&	11.90	&	12.70	&	13.49	&	14.77	&	Sch	&	FLI	\\
05.04.2016	&	57483.519	&	11.96	&	12.76	&	13.62	&	14.81	&	2-m	&	VA	\\
06.04.2016	&	57484.522	&	12.00	&	12.79	&	13.60	&	14.85	&	Sch	&	FLI	\\
06.04.2016	&	57484.559	&	12.09	&	12.91	&	13.77	&	15.07	&	2-m	&	VA	\\
27.04.2016	&	57506.462	&	11.92	&	12.73	&	13.53	&	14.76	&	Sch	&	FLI	\\
13.05.2016	&	57522.425	&	12.02	&	12.84	&	13.64	&	14.86	&	Sch	&	FLI	\\
14.05.2016	&	57523.426	&	12.01	&	12.85	&	13.68	&	14.91	&	Sch	&	FLI	\\
31.05.2016	&	57540.421	&	12.17	&	12.99	&	13.87	&	15.01	&	2-m	&	VA	\\
25.06.2016	&	57565.421	&	12.05	&	12.87	&	13.71	&	14.95	&	Sch	&	FLI	\\
02.07.2016	&	57572.341	&	12.03	&	12.83	&	13.66	&	14.85	&	2-m	&	VA	\\
11.07.2016	&	57581.382	&	11.96	&	12.73	&	13.54	&	14.72	&	Sch	&	FLI	\\
12.07.2016	&	57582.418	&	11.94	&	12.74	&	13.54	&	14.74	&	Sch	&	FLI	\\
13.07.2016	&	57583.397	&	11.99	&	12.79	&	13.60	&	14.84	&	Sch	&	FLI	\\
01.08.2016	&	57602.381	&	12.30	&	13.14	&	14.03	&	15.36	&	2-m	&	VA	\\
02.08.2016	&	57603.365	&	12.25	&	13.08	&	13.97	&	15.26	&	2-m	&	VA	\\
04.08.2016	&	57605.389	&	12.15	&	13.02	&	13.89	&	15.14	&	Sch	&	FLI	\\
05.08.2016	&	57606.380	&	12.17	&	13.05	&	13.92	&	15.20	&	Sch	&	FLI	\\
06.08.2016	&	57607.369	&	12.17	&	13.03	&	13.89	&	15.15	&	Sch	&	FLI	\\
11.09.2016	&	57643.305	&	12.21	&	13.05	&	13.90	&	15.18	&	Sch	&	FLI	\\
02.10.2016	&	57664.268	&	11.98	&	12.79	&	13.60	&	14.82	&	Sch	&	FLI	\\
05.11.2016	&	57698.239	&	11.99	&	12.82	&	13.63	&	14.86	&	Sch	&	FLI	\\
21.11.2016	&	57714.251	&	12.36	&	13.26	&	14.19	&	15.51	&	2-m	&	VA	\\
22.11.2016	&	57715.226	&	12.33	&	13.26	&	14.24	&	15.56	&	2-m	&	VA	\\
23.11.2016	&	57716.244	&	12.31	&	13.20	&	14.14	&	15.45	&	2-m	&	VA	\\
01.01.2017	&	57755.210	&	12.15	&	13.04	&	13.92	&	15.15	&	Sch	&	FLI	\\
02.01.2017	&	57756.222	&	12.08	&	12.93	&	13.80	&	15.07	&	Sch	&	FLI	\\
28.01.2017	&	57782.194	&	12.09	&	12.91	&	13.83	&	15.14	&	2-m	&	VA	\\
30.01.2017	&	57784.200	&	12.09	&	12.94	&	13.79	&	14.95	&	2-m	&	VA	\\
31.01.2017	&	57785.197	&	12.21	&	13.04	&	13.91	&	15.12	&	2-m	&	VA	\\
01.02.2017	&	57786.197	&	12.28	&	13.15	&	13.99	&	15.25	&	2-m	&	VA	\\
17.02.2017	&	57801.620	&	12.10	&	12.94	&	13.77	&	15.02	&	Sch	&	FLI	\\
05.03.2017	&	57817.573	&	11.99	&	12.80	&	13.62	&	14.85	&	Sch	&	FLI	\\
02.04.2017	&	57845.559	&	12.06	&	12.88	&	13.70	&	14.93	&	Sch	&	FLI	\\
03.04.2017	&	57846.587	&	12.07	&	12.90	&	13.72	&	14.97	&	Sch	&	FLI	\\
02.05.2017	&	57875.505	&	12.02	&	12.81	&	13.61	&	14.84	&	2-m	&	VA	\\
18.05.2017	&	57892.396	&	12.11	&	12.96	&	13.77	&	15.03	&	Sch	&	FLI	\\
19.05.2017	&	57893.405	&	12.02	&	12.81	&	13.62	&	14.79	&	2-m	&	VA	\\
30.05.2017	&	57904.409	&	12.01	&	12.81	&	13.64	&	14.86	&	Sch	&	FLI	\\
01.08.2017	&	57967.412	&	11.93	&	12.70	&	13.49	&	14.69	&	Sch	&	FLI	\\
02.08.2017	&	57968.298	&	11.93	&	12.72	&	13.51	&	14.71	&	Sch	&	FLI	\\
03.08.2017	&	57969.297	&	11.94	&	12.71	&	13.52	&	14.73	&	Sch	&	FLI	\\
12.08.2017	&	57978.469	&	11.91	&	12.68	&	13.47	&	14.64	&	Sch	&	FLI	\\
14.09.2017	&	58011.295	&	11.92	&	12.71	&	13.50	&	14.69	&	Sch	&	FLI	\\
15.09.2017	&	58012.313	&	11.92	&	12.69	&	13.48	&	14.68	&	Sch	&	FLI	\\
16.09.2017	&	58013.300	&	11.90	&	12.68	&	13.47	&	14.67	&	Sch	&	FLI	\\
12.10.2017	&	58039.283	&	11.93	&	12.72	&	13.50	&	14.71	&	Sch	&	FLI	\\
14.10.2017	&	58041.278	&	11.91	&	12.67	&	13.48	&	-    	&	2-m	&	VA	\\
16.10.2017	&	58043.265	&	11.87	&	12.65	&	13.45	&	14.63	&	Sch	&	FLI	\\
16.10.2017	&	58043.297	&	11.91	&	12.68	&	13.52	&	14.50	&	2-m	&	VA	\\
17.10.2017	&	58044.324	&	11.88	&	12.66	&	13.46	&	14.63	&	Sch	&	FLI	\\
18.10.2017	&	58045.371	&	11.88	&	12.69	&	13.49	&	14.68	&	Sch	&	FLI	\\
22.11.2017	&	58080.267	&	12.25	&	13.16	&	14.04	&	15.33	&	Sch	&	FLI	\\
23.11.2017	&	58081.275	&	12.22	&	13.12	&	13.97	&	15.27	&	Sch	&	FLI	\\
21.12.2017	&	58109.297	&	12.41	&	13.35	&	14.24	&	15.61	&	Sch	&	FLI	\\
25.12.2017	&	58113.215	&	12.27	&	13.19	&	14.08	&	15.40	&	Sch	&	FLI	\\
26.12.2017	&	58114.249	&	12.26	&	13.17	&	14.09	&	15.42	&	Sch	&	FLI	\\
09.04.2018	&	58217.593	&	12.04	&	12.87	&	13.68	&	14.94	&	Sch	&	FLI	\\
10.04.2018	&	58218.580	&	12.04	&	12.87	&	13.67	&	14.91	&	Sch	&	FLI	\\
12.04.2018	&	58220.536	&	12.09	&	12.94	&	13.78	&	15.03	&	Sch	&	FLI	\\
08.06.2018	&	58278.413	&	12.12	&	12.98	&	13.83	&	15.10	&	Sch	&	FLI	\\
11.06.2018	&	58281.451	&	12.11	&	12.95	&	13.85	&	15.08	&	2-m	&	AND	\\
12.07.2018	&	58312.405	&	12.06	&	12.92	&	13.76	&	15.03	&	Sch	&	FLI	\\
16.07.2018	&	58316.366	&	12.17	&	13.05	&	13.92	&	15.23	&	Sch	&	FLI	\\
09.08.2018	&	58340.356	&	12.02	&	12.90	&	13.74	&	15.00	&	Sch	&	FLI	\\
11.08.2018	&	58342.355	&	12.02	&	12.86	&	13.71	&	14.96	&	Sch	&	FLI	\\
12.08.2018	&	58343.367	&	12.02	&	12.87	&	13.70	&	14.93	&	Sch	&	FLI	\\
13.08.2018	&	58344.347	&	12.06	&	12.93	&	13.77	&	15.06	&	Sch	&	FLI	\\
15.08.2018	&	58346.299	&	12.04	&	12.87	&	13.74	&	14.92	&	2-m	&	AND	\\
01.09.2018	&	58363.321	&	12.07	&	12.91	&	13.76	&	15.02	&	Sch	&	FLI	\\
02.09.2018	&	58364.300	&	12.00	&	12.83	&	13.67	&	14.91	&	Sch	&	FLI	\\
17.10.2018	&	58409.266	&	12.02	&	12.88	&	13.71	&	14.99	&	Sch	&	FLI	\\
05.11.2018	&	58428.255	&	11.96	&	12.76	&	13.57	&	14.77	&	Sch	&	FLI	\\
12.11.2018	&	58435.234	&	11.93	&	12.75	&	13.54	&	14.74	&	Sch	&	FLI	\\
08.01.2019	&	58492.200	&	11.98	&	12.77	&	13.60	&	14.82	&	Sch	&	FLI	\\
05.03.2019	&	58547.601	&	12.00	&	12.82	&	13.63	&	14.85	&	Sch	&	FLI	\\
29.04.2019	& 58603.457	& 12.04	& 12.85	& 13.66	& 14.91	& Sch	& FLI \\
01.05.2019	& 58604.510	& 12.02	& 12.84	& 13.66	& 14.88	& Sch	& FLI \\
02.05.2019	& 58606.428	& 12.04	& 12.85	& 13.72	& 15.01	& 2-m	& AND \\
03.05.2019	& 58606.530	& 12.03	& 12.87	& 13.69	& 14.93	& Sch	& FLI \\
30.06.2019	& 58665.392	& 12.10	& 12.96	& 13.79	& 15.05	& Sch	& FLI \\
01.07.2019	& 58666.410	& 12.24	& 13.15	& 14.02	& 15.33	& Sch	& FLI \\
02.07.2019	& 58667.396	& 12.34	& 13.28	& 14.17	& 15.51	& Sch	& FLI \\
\hline \hline
\end{longtable}}

\subsection*{3.2. FHO 27}

FHO 27 was included in the list of young stellar object candidates in the work of Guieu et al. (2009).
In the work of Findeisen et al. (2013) the $R$ band light curve of the star for the period 2009$-$2012 was presented.
The authors reported that FHO 27 exhibiting multiple fading events.

The results from our $BVRI$ observations of FHO 27 have summarized in Tab. 2\footnote{The photometric data are available also at CDS (\url{http://cds.u-strasbg.fr/}}).
The columns have the same content as in Tab. 1.
The photometric limit of our data obtained with the 50/70-cm Schmidt telescope is about 19.5 mag.
Therefore, the lowest states of the star in the $V$ and $B$ bands below this limit are not registered.
The received new photometric data of FHO 27 and the data from Ibryamov et al. (2015a) are presented in Fig. 5.

\begin{figure*}
   \centering
   \includegraphics[width=11cm, angle=0]{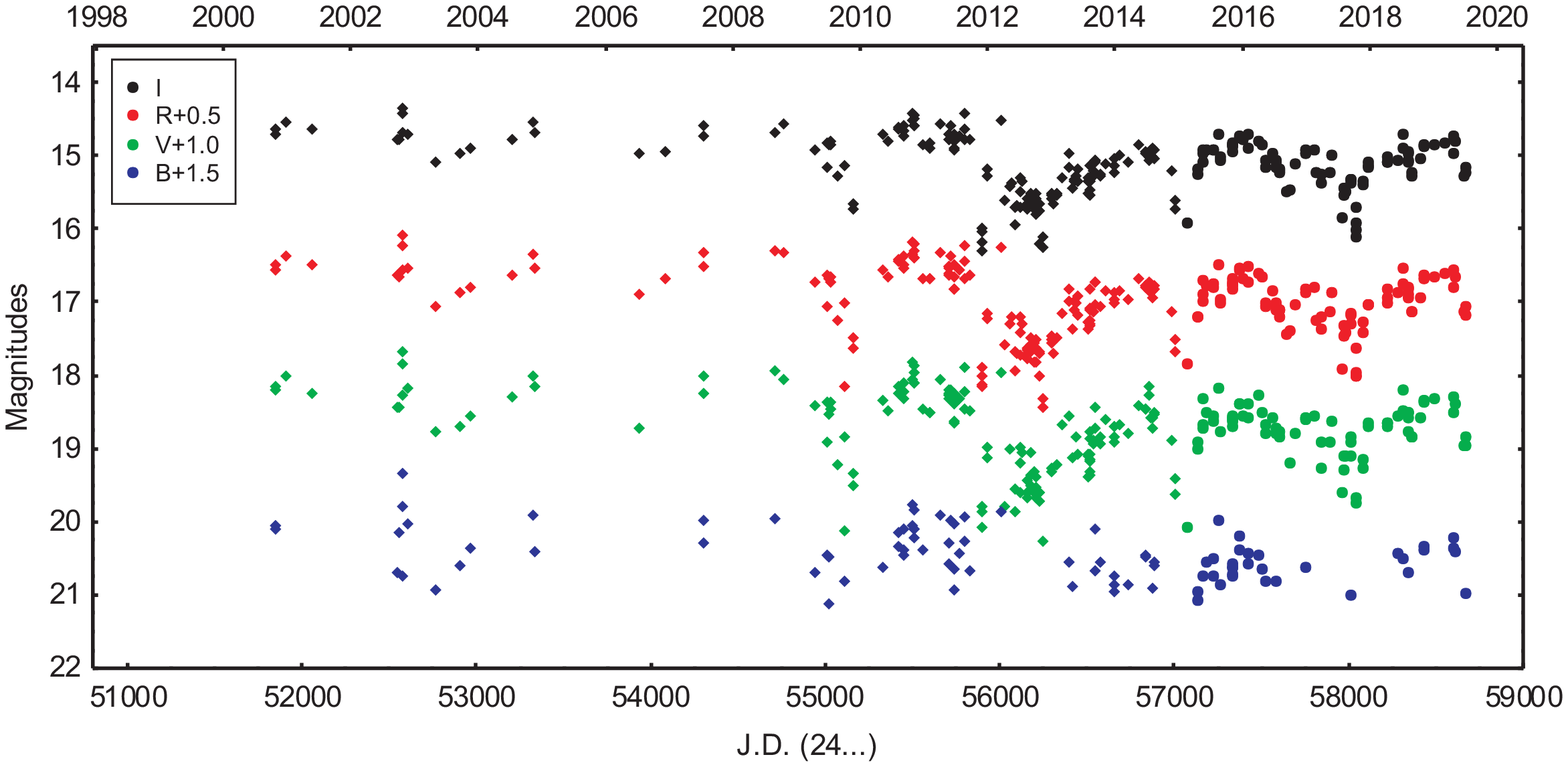}
   \caption{$BVRI$ light curves of FHO 27 Cyg for the period October 2000$-$July 2019. The circles denote the new photometric data (present paper), and the diamonds represent the data from Ibryamov et al. (2015a). The registered amplitudes of the star's variability during the new observations are 1.39 mag in the $I$ band, 1.54 mag in the $R$ band, $>$1.91 mag in the $V$ band, and $>$1.09 in the $B$ band.}\label{Fig5}
   \end{figure*}

It can be seen from Fig. 5 that during the period February 2015$-$July 2019 several fading events in the light curves of FHO 27 were registered.
The deepest of them were observed at the beginning of 2015 and in October 2017.
After 2012 there is a slight decrease in the registered maximum levels of the star's brightness.
The possible reason for this phenomenon is a decrease in the accretion rate or increased circumstellar extinction.

{\footnotesize
\begin{longtable}{cccccccc}
\caption{$BVRI$ photometry of FHO 27}\\
\hline\hline
\noalign{\smallskip}  
Date \hspace{1.0cm} &	J.D. (24...) \hspace{1mm}	&	$I$ [mag]\hspace{2mm} & $R$ [mag]\hspace{2mm} & $V$ [mag]\hspace{2mm} & $B$ [mag]\hspace{2mm} & Tel \hspace{1.0mm} & CCD\\
\noalign{\smallskip}  
\hline
\endfirsthead
\caption{continued.}\\
\hline\hline
\noalign{\smallskip}  
Date \hspace{1.0cm} &	J.D. (24...) \hspace{1mm}	&	$I$ [mag]\hspace{2mm} & $R$ [mag]\hspace{2mm} & $V$ [mag]\hspace{2mm} & $B$ [mag]\hspace{2mm} & Tel \hspace{1.0mm} & CCD\\
\noalign{\smallskip}  
\hline
\noalign{\smallskip}  
\endhead
\hline
\label{Tab2}
\endfoot
\noalign{\smallskip}
21.02.2015	&	57074.641	&	15.92	&	17.35	&	19.08	&	-	&	Sch	&	FLI	\\
23.04.2015	&	57136.487	&	15.19	&	16.69	&	17.92	&	19.46	&	Sch	&	FLI	\\
25.04.2015	&	57138.491	&	15.26	&	16.69	&	18.00	&	19.58	&	Sch	&	FLI	\\
18.05.2015	&	57161.423	&	14.92	&	16.21	&	17.32	&	-	&	Sch	&	FLI	\\
19.05.2015	&	57162.460	&	15.10	&	16.48	&	17.72	&	19.24	&	Sch	&	FLI	\\
21.05.2015	&	57164.459	&	14.97	&	16.39	&	17.68	&	-	&	Sch	&	FLI	\\
12.06.2015	&	57186.444	&	14.92	&	16.28	&	17.52	&	19.04	&	Sch	&	FLI	\\
16.07.2015	&	57220.364	&	14.92	&	16.25	&	17.55	&	19.00	&	Sch	&	FLI	\\
17.17.2015	&	57221.417	&	14.92	&	16.29	&	17.63	&	19.23	&	Sch	&	FLI	\\
24.08.2015	&	57259.337	&	14.72	&	15.98	&	17.17	&	18.49	&	Sch	&	FLI	\\
25.08.2015	&	57260.342	&	15.07	&	16.51	&	-	&	-	&	Sch	&	FLI	\\
03.09.2015	&	57269.328	&	15.02	&	16.46	&	17.77	&	19.37	&	Sch	&	FLI	\\
03.11.2015	&	57330.191	&	14.88	&	16.24	&	17.60	&	19.12	&	Sch	&	FLI	\\
04.11.2015	&	57331.230	&	14.95	&	16.33	&	17.69	&	19.22	&	Sch	&	FLI	\\
05.11.2015	&	57332.243	&	14.84	&	16.27	&	17.56	&	19.23	&	Sch	&	FLI	\\
06.11.2015	&	57333.228	&	14.86	&	16.23	&	17.63	&	19.07	&	Sch	&	FLI	\\
07.11.2015	&	57334.219	&	14.82	&	16.19	&	17.57	&	19.13	&	Sch	&	FLI	\\
15.12.2015	&	57372.216	&	14.73	&	16.04	&	17.40	&	18.69	&	Sch	&	FLI	\\
17.12.2015	&	57374.269	&	14.75	&	16.07	&	17.38	&	18.89	&	Sch	&	FLI	\\
02.01.2016	&	57390.194	&	14.77	&	16.19	&	17.56	&	-	&	Sch	&	FLI	\\
06.02.2016	&	57425.204	&	14.72	&	16.02	&	17.40	&	18.93	&	Sch	&	FLI	\\
07.02.2016	&	57426.200	&	14.90	&	16.23	&	17.59	&	19.08	&	Sch	&	FLI	\\
06.04.2016	&	57484.522	&	14.81	&	16.11	&	17.27	&	18.95	&	Sch	&	FLI	\\
27.04.2016	&	57506.462	&	14.84	&	16.16	&	17.51	&	19.14	&	Sch	&	FLI	\\
13.05.2016	&	57522.425	&	15.07	&	16.56	&	17.80	&	19.31	&	Sch	&	FLI	\\
14.05.2016	&	57523.426	&	15.15	&	16.52	&	17.68	&	-	&	Sch	&	FLI	\\
25.06.2016	&	57565.421	&	14.98	&	16.35	&	17.57	&	-	&	Sch	&	FLI	\\
11.07.2016	&	57581.382	&	15.10	&	16.57	&	17.80	&	19.31	&	Sch	&	FLI	\\
12.07.2016	&	57582.418	&	15.07	&	16.50	&	17.73	&	-	&	Sch	&	FLI	\\
13.07.2016	&	57583.397	&	15.16	&	16.60	&	17.77	&	-	&	Sch	&	FLI	\\
04.08.2016	&	57605.389	&	15.18	&	16.60	&	17.77	&	-	&	Sch	&	FLI	\\
05.08.2016	&	57606.380	&	15.20	&	16.69	&	17.84	&	-	&	Sch	&	FLI	\\
06.08.2016	&	57607.369	&	15.23	&	16.60	&	-	&	-	&	Sch	&	FLI	\\
11.09.2016	&	57643.305	&	15.50	&	16.95	&	-	&	-	&	Sch	&	FLI	\\
02.10.2016	&	57664.268	&	15.46	&	16.89	&	18.19	&	-	&	Sch	&	FLI	\\
05.11.2016	&	57698.239	&	15.12	&	16.54	&	17.80	&	-	&	Sch	&	FLI	\\
01.01.2017	&	57755.210	&	14.92	&	16.32	&	17.57	&	-	&	Sch	&	FLI	\\
02.01.2017	&	57756.222	&	14.97	&	16.37	&	17.61	&	19.11	&	Sch	&	FLI	\\
17.02.2017	&	57801.620	&	14.93	&	16.30	&	17.55	&	-	&	Sch	&	FLI	\\
05.03.2017	&	57817.573	&	15.24	&	16.76	&	-	&	-	&	Sch	&	FLI	\\
02.04.2017	&	57845.559	&	15.38	&	16.86	&	18.26	&	-	&	Sch	&	FLI	\\
03.04.2017	&	57846.587	&	15.26	&	16.71	&	17.92	&	-	&	Sch	&	FLI	\\
18.05.2017	&	57892.396	&	15.24	&	16.64	&	17.92	&	-	&	Sch	&	FLI	\\
30.05.2017	&	57904.409	&	15.00	&	16.37	&	17.62	&	-	&	Sch	&	FLI	\\
01.08.2017	&	57967.412	&	15.84	&	17.42	&	18.60	&	-	&	Sch	&	FLI	\\
02.08.2017	&	57968.298	&	15.54	&	16.96	&	18.28	&	-	&	Sch	&	FLI	\\
03.08.2017	&	57969.297	&	15.44	&	16.83	&	18.09	&	-	&	Sch	&	FLI	\\
12.08.2017	&	57978.469	&	15.49	&	16.91	&	18.11	&	-	&	Sch	&	FLI	\\
14.09.2017	&	58011.295	&	15.33	&	16.66	&	17.84	&	19.50	&	Sch	&	FLI	\\
15.09.2017	&	58012.313	&	15.37	&	16.79	&	18.09	&	-	&	Sch	&	FLI	\\
16.09.2017	&	58013.300	&	15.32	&	16.67	&	17.92	&	-	&	Sch	&	FLI	\\
12.10.2017	&	58039.283	&	15.71	&	17.13	&	-	&	-	&	Sch	&	FLI	\\
16.10.2017	&	58043.265	&	15.93	&	17.46	&	18.67	&	-	&	Sch	&	FLI	\\
17.10.2017	&	58044.324	&	16.11	&	17.51	&	18.75	&	-	&	Sch	&	FLI	\\
18.10.2017	&	58045.371	&	16.02	&	17.52	&	-	&	-	&	Sch	&	FLI	\\
22.11.2017	&	58080.267	&	15.35	&	16.78	&	18.16	&	-	&	Sch	&	FLI	\\
23.11.2017	&	58081.275	&	15.39	&	16.91	&	18.27	&	-	&	Sch	&	FLI	\\
21.12.2017	&	58109.297	&	15.09	&	16.53	&	17.65	&	-	&	Sch	&	FLI	\\
25.12.2017	&	58113.215	&	15.14	&	16.54	&	17.71	&	-	&	Sch	&	FLI	\\
26.12.2017	&	58114.249	&	15.17	&	16.54	&	-	&	-	&	Sch	&	FLI	\\
09.04.2018	&	58217.593	&	15.10	&	16.52	&	17.66	&	-	&	Sch	&	FLI	\\
10.04.2018	&	58218.580	&	15.10	&	16.45	&	17.70	&	-	&	Sch	&	FLI	\\
12.04.2018	&	58220.536	&	15.01	&	16.31	&	17.66	&	-	&	Sch	&	FLI	\\
08.06.2018	&	58278.413	&	15.06	&	16.36	&	17.55	&	18.92	&	Sch	&	FLI	\\
12.07.2018	&	58312.405	&	14.72	&	16.04	&	17.20	&	-	&	Sch	&	FLI	\\
16.07.2018	&	58316.366	&	14.91	&	16.26	&	17.49	&	19.01	&	Sch	&	FLI	\\
09.08.2018	&	58340.356	&	15.07	&	16.40	&	17.59	&	-	&	Sch	&	FLI	\\
11.08.2018	&	58342.355	&	15.08	&	16.45	&	17.77	&	-	&	Sch	&	FLI	\\
12.08.2018	&	58343.367	&	14.94	&	16.29	&	17.51	&	-	&	Sch	&	FLI	\\
13.08.2018	&	58344.347	&	14.97	&	16.33	&	17.51	&	19.19	&	Sch	&	FLI	\\
01.09.2018	&	58363.321	&	15.24	&	16.64	&	17.85	&	-	&	Sch	&	FLI	\\
02.09.2018	&	58364.300	&	15.27	&	16.62	&	17.85	&	-	&	Sch	&	FLI	\\
17.10.2018	&	58409.266	&	15.05	&	16.44	&	17.57	&	-	&	Sch	&	FLI	\\
05.11.2018	&	58428.255	&	14.85	&	16.14	&	17.36	&	18.89	&	Sch	&	FLI	\\
12.11.2018	&	58435.234	&	14.87	&	16.17	&	17.35	&	18.84	&	Sch	&	FLI	\\
08.01.2019	&	58492.200	&	14.84	&	16.15	&	17.32	&	-	&	Sch	&	FLI	\\
05.03.2019	&	58547.601	&	14.82	&	16.11	&	-	&	-	&	Sch	&	FLI	\\
29.04.2019	& 58603.457	& 14.98	& 16.29	& 17.50	& 18.86	& Sch	& FLI \\
01.05.2019	& 58604.510	& 14.74	& 16.06	& 17.30	& 18.72	& Sch	& FLI \\
03.05.2019	& 58606.530	& 14.81	& 16.15	& 17.39	& 18.90	& Sch	& FLI \\
30.06.2019	& 58665.392	& 15.28	& 16.63	& 17.95	&	- & Sch	& FLI \\
01.07.2019	& 58666.410	& 15.24	& 16.67	& 17.95	& - &	Sch	& FLI \\
02.07.2019	& 58667.396	& 15.16	& 16.55	& 17.83	& 19.47	& Sch	& FLI \\
\hline \hline
\end{longtable}}

The measured color indices versus the stellar $V$ magnitude of FHO 27 are plotted in Fig. 6.
It can be seen that initially, the star becomes redder when fainter.
Such color variations are typical for PMS stars, whose variability is produced by obscuration by the circumstellar material.
The figure shows evidence for a blueing effect during the minima in the star's brightness.
This trend is the best seen in the $B-V$ index.
In this case, the ratio of scattered light to direct light increases and the star's color turns bluer.

\begin{figure}[h!]
\begin{center}
\includegraphics[width=7.5cm]{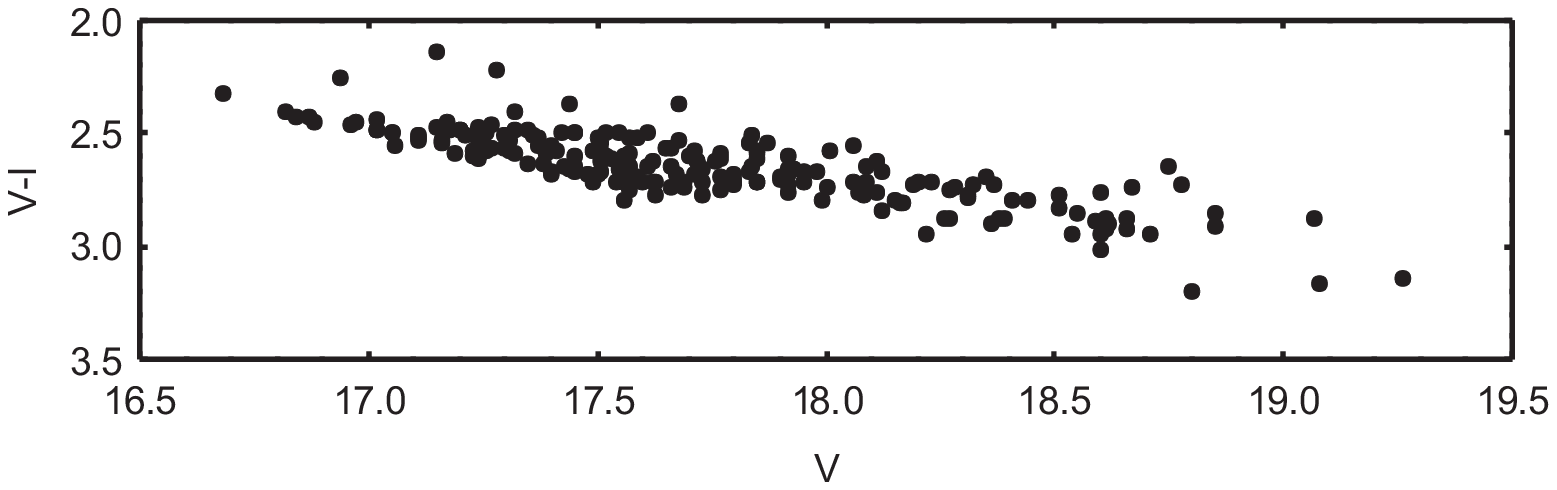}
\includegraphics[width=7.5cm]{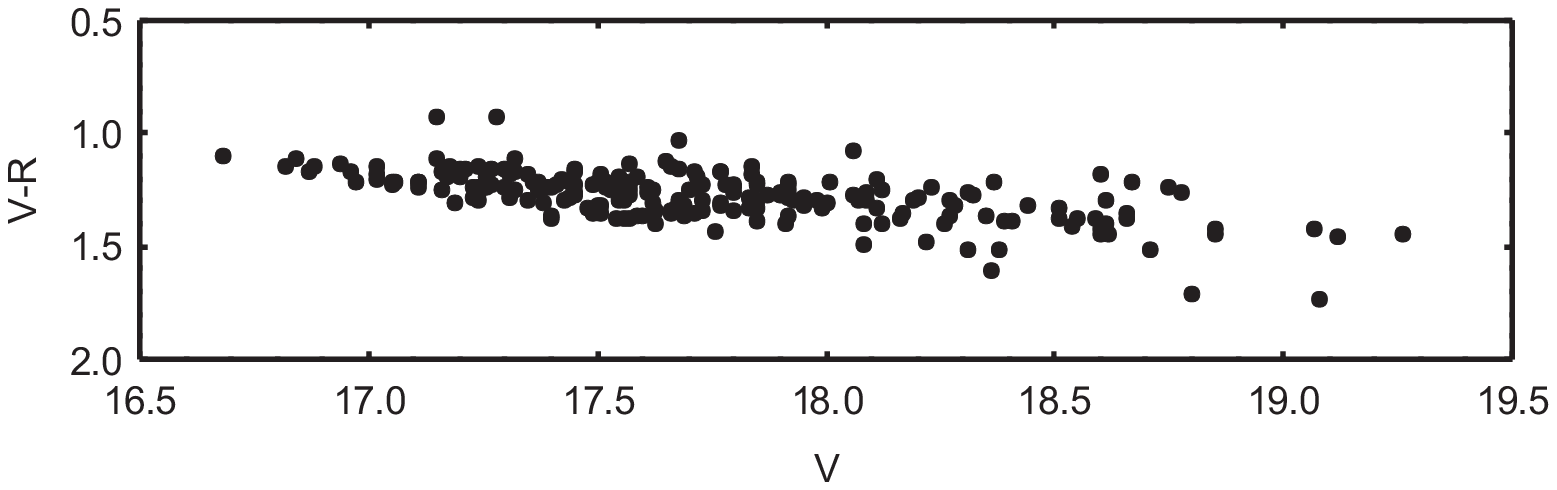}
\includegraphics[width=7.5cm]{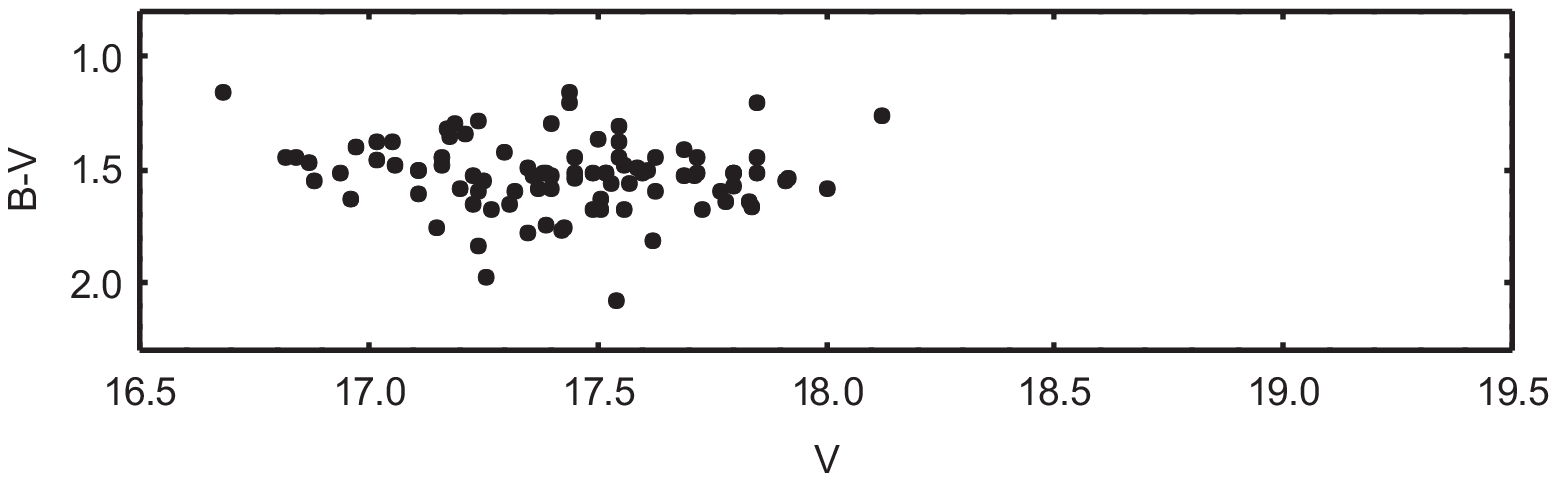}
\caption{Color indices $V-I$, $V-R$, and $B-V$ versus the stellar $V$ magnitude of FHO 27.}\label{Fig6}
\end{center}
\end{figure}

An important result of our study of FHO 27 is the discovery of periodicity in its variability.
The period is available using the photometric data collected after January 2012.
The obtained periodogram of the star using the data in the $R$ band is shown in Fig. 7 (left).
A significant peak corresponding to 893 days was registered.
The $R$ band phased light curve of the star according to the discovered period is plotted in Fig. 7 (right).
The possible reasons for the periodicity in the brightness variations of FHO 27 are the same as in V521 Cyg.

\begin{figure}[h!]
\begin{center}
\includegraphics[width=6cm, angle=0]{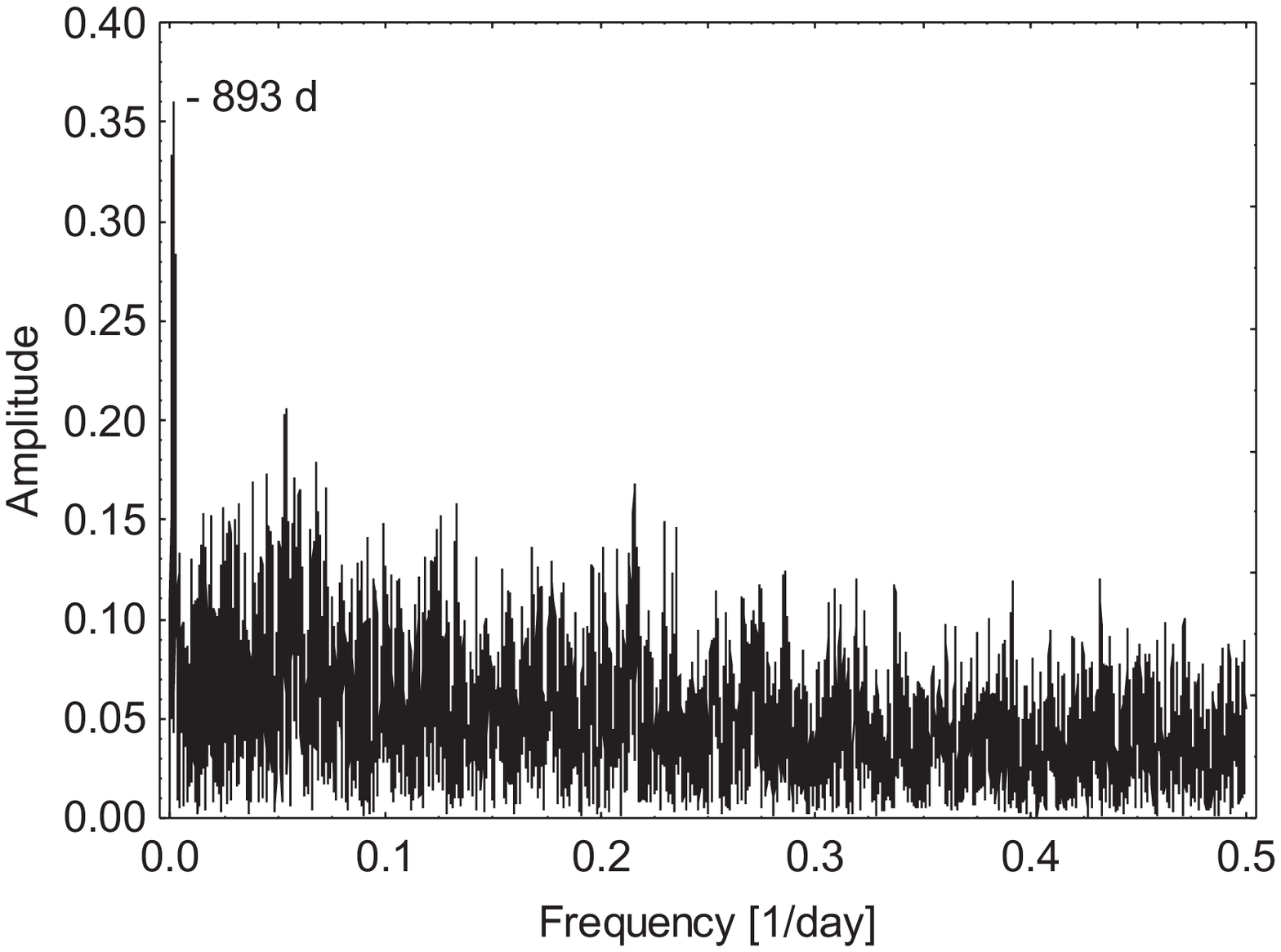}
\includegraphics[width=6cm, angle=0]{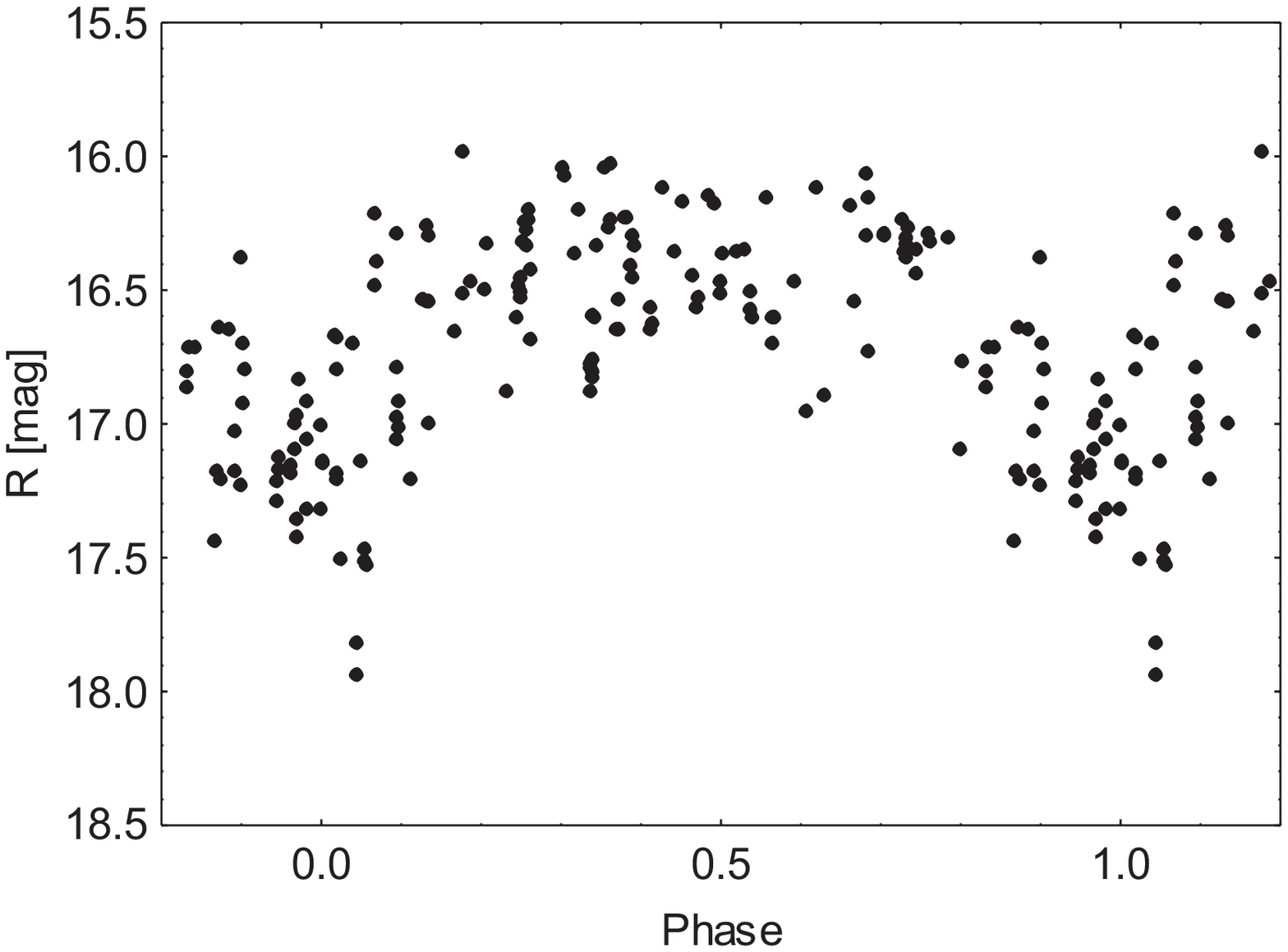}
\caption{Left: Periodogram analysis of the photometric data of FHO 27. Right: $R$ band phased light curve of FHO 27.}\label{Fig7}
\end{center}
\end{figure}

\section*{ 4. Concluding remarks}

We presented and discussed the $BVRI$ observations of the young stellar objects V521 Cyg and FHO 27 obtained in the period 2015$-$2019.
They show that the stars exhibit strong optical variability and deep fading events in the brightness.
Using the new data and the ones available in the literature, we built the long-term $BVRI$ light curves of the stars.
We specified 594 days periodicity in the photometric behavior of V521 Cyg, and we found an 893 days periodicity in the brightness variations of FHO 27.
We plan to continue our four-color observational monitoring of the field of LDN 935 during the next years.

\section*{Acknowledgements}

This work is supported by the Bulgarian Ministry of Education and Science under the National Program for Research "Young Scientists and Postdoctoral Students".
This research has made use of NASA's Astrophysics Data System.
We thank the anonymous referee for carefully reading the text and for the useful suggestions and comments that helped to improve the paper.


\begin{thebibliography}{}

\bibitem{}
Armond, T., Reipurth, B., Bally, J. \& Aspin, C., 2011, {\em A\&A, 528, A125}

\bibitem{}
Audard, M., {\'A}brah{\'a}m, P., Dunham, M. M. et al., 2014, {\em in Protostars and Planets VI, eds. H. Beuther, R. S. Klessen, C. P. Dullemond \& T. Henning (Tucson: University of Arizona Press), 387}

\bibitem{}
Bally, J., Ginsburg, A., Probst, R., Reipurth, B., Shirley, Y. L. \& Stringfellow, G. S., 2014, {\em AJ, 148, 120}

\bibitem{}
Bhardwaj, A., Panwar, N., Herczeg, G. J., Chen, W. P. \& Singh, H. P., 2019, {\em A\&A, 627, A135}

\bibitem{}
Bibo, E. A. \& Th{\'e}, P. S., 1990, {\em A\&A, 236, 155}

\bibitem{}
Cody, A. M., Stauffer, J., Baglin, A. et al., 2014, {\em AJ, 147, 82}

\bibitem{}
Dullemond, C. P., van den Ancker, M. E., Acke, B. \& van Boekel, R., 2003, {\em ApJ, 594, L47-L50}

\bibitem{}
Erastova, L. K. \& Tsvetkov, M. K., 1978, {\em Perem. Zvezdy, 21, 79}

\bibitem{}
Fernie, J. D., 1983, {\em PASP, 95, 782}

\bibitem{}
Filin, A. Ya., 1974, {\em Perem. Zvezdy Prilozh., 2, 63}

\bibitem{}
Findeisen, K., Hillenbrand, L., Ofek, E., Levitan, D., Sesar, B., Laher, R. \& Surace, J., 2013, {\em ApJ, 768, 93}

\bibitem{}
Froebrich, D., Scholz, A., Campbell-White, J. et al., 2018, {\em RNAAS, 2, 61}

\bibitem{}
Giannini, T., Munari, U., Antoniucci, S., Lorenzetti, D., Arkharov, A. A., Dallaporta, S., Rossi, A. \& Traven, G., 2018, {\em A\&A, 611, A54}

\bibitem{}
Grankin, K. N., Melnikov, S. Yu., Bouvier, J., Herbst, W. \& Shevchenko, V. S., 2007, {\em A\&A, 461, 183}

\bibitem{}
Grinin, V. P., Kiselev, N. N., Minikulov, N. Kh., Chernova, G. P. \& Voshchinnikov, N. V., 1991, {\em Ap\&SS, 186, 283}

\bibitem{}
Guieu, S., Rebull, L. M., Stauffer, J. R. et al., 2009, {\em ApJ, 697, 787}

\bibitem{}
Herbig, G. H., 1958, {\em ApJ, 128, 259}

\bibitem{}
Herbst, W., Herbst, D. K., Grossman, E. J. \& Weinstein, D., 1994, {\em AJ, 108, 1906}

\bibitem{}
Herbst, W. \& Shevchenko, V. S., 1999, {\em AJ, 118, 1043}

\bibitem{}
Hillenbrand, L. A., Strom, S. E., Vrba, F. J. \& Keene, J., 1992, {\em ApJ, 397, 613}

\bibitem{}
Hoffmeister, C., 1949, {\em AN, 278, 24}

\bibitem{}
Ibryamov, S. I., Semkov, E. H. \& Peneva, S. P., 2015a, {\em PASA, 32, e021}

\bibitem{}
Ibryamov, S. I., Semkov, E. H. \& Peneva, S. P., 2015b, {\em Bulgarian Astronomical Journal, 22, 3}

\bibitem{}
Ibryamov, S. I. \& Semkov, E. H., 2016, {\em Bulgarian Astronomical Journal, 24, 62}

\bibitem{}
Ibryamov, S. I., Semkov, E. H. \& Peneva, S. P., 2018a, {\em PASA, 35, e007}

\bibitem{}
Ibryamov, S., Semkov, E., Milanov, T. \& Peneva, S., 2018b, {\em RAA, 18, 137}

\bibitem{}
Lenz, P. \& Breger, M., 2005, {\em Communications in Asteroseismology, 146, 53}

\bibitem{}
Lynds, B. T., 1962, {\em ApJS, 7, 1}

\bibitem{}
Mitskevich, M. P. \& Pavlenko, E. P., 2001, {\em Astrophysics, 44, 411}

\bibitem{}
Natta, A. \& Whitney, B. A., 2000, {\em A\&A, 364, 633}

\bibitem{}
Petrov, P. P., 2003, {\em Astrophysics, 46, 506}

\bibitem{}
Poljan\v{c}i\'{c} Beljan, I., Jurdana-\v{S}epi\'{c}, R., Semkov, E. H., Ibryamov, S., Peneva, S. P. \& Tsvetkov, M. K., 2014, {\em A\&A, 568, A49}

\bibitem{}
Reipurth, B. \& Aspin, C., 2010, {\em in Evolution of Cosmic Objects through their Physical Activity, eds. H. A. Harutyunian, A. M. Mickaelian \& Y. Terzian (Yerevan: Gitutyun), 19}

\bibitem{}
Semkov, E. H., Peneva, S. P., Munari, U., Milani, A. \& Valisa, P., 2010, {\em A\&A, 523, L3}

\bibitem{}
Semkov, E. H., Peneva, S. P. \& Ibryamov, S. I., 2017, {\em Bulgarian Astronomical Journal, 26, 57}

\bibitem{}
Voshchinnikov, N. V., 1989, {\em Astrofizika, 30, 509}

\end{thebibliography}
\end{document}